\documentclass[%
 reprint,
 amsmath,amssymb,
 aps,
]{revtex4-2}

\usepackage{graphicx}
\usepackage{dcolumn}
\usepackage{bm}

\usepackage[T1]{fontenc} 
\usepackage{tabstackengine}
\setstackEOL{\cr}
\usepackage{hyperref}
\usepackage{soul}
\usepackage{amsmath}
\usepackage{mathtools}
\usepackage[italic]{hepnames}
\usepackage{graphicx}
\usepackage{float}

\newcommand*{\ctZ}{\ensuremath{c_{tZ}}\xspace}
\newcommand*{\tWZ}{\ensuremath{tWZ}\xspace}
\newcommand*{\tZq}{\Pqt{}\PZ{}\Pq}

\newcommand*{\ttZ}{\ensuremath{\Pqt{}\Paqt{}\PZ{}}\xspace}
\newcommand*{\WZbb}{\ensuremath{\PW{}\PZ{}\Paqb{}\Paqb{}}\xspace}
\newcommand*{\ZZjj}{\ensuremath{\PZ{}\PZ{}jj}\xspace}
\newcommand{\pt}{\ensuremath{p_{T}}\xspace}
\newcommand{\ptz}{\ensuremath{p^{Z}_{T}}\xspace}

\newcommand{\dsigptz}{\ensuremath{\frac{d \sigma_{\tWZ}}{d \ptz}}\xspace}

\newcommand{\ci}{\ensuremath{c_{i}}\xspace}
\newcommand{\priorfunc}{\ensuremath{p(c_{i})}\xspace}
\newcommand{\postfunc}{\ensuremath{p(c_{i}| x,\Sigma)}\xspace}
\newcommand{\likelihood}{(\ensuremath{d(c_{i}, x)^{T} \Sigma^{-1}d(c_{i}, x))} \xspace}

\def\mg{\textsc{MadGraph5\_aMC@NLO}\xspace}
\def\mgversion{\textsc{v2\_6\_7}\xspace}
\def\pythia{\textsc{PYTHIA}\xspace}
\def\pythiaversion{\textsc{v8.306}\xspace}
\def\delphes{\textsc{DELPHES}\xspace}
\def\delphesversion{\textsc{v3\_5\_0}\xspace}
\def\smeftatnlo{\textsc{SMEFTatNLO}\xspace}
\def\roounfold{\textsc{ROOUNFOLD}\xspace}

\def\mc{\textsf{emcee}\xspace}
\def\corner{\textsf{corner}\xspace}
\def\deft{\textsf{\emph{d}EFT}\xspace}
\def\smefit{\textsf{SMEFiT}\xspace}
\def\ifit{\textsf{IFIT}\xspace}

\newcommand*{\deftlink}{\href{https://github.com/keaveney/dEFT/releases/tag/tWZ_HLLHC_PRD_resub}{\deft on GitHub\xspace}}

\begin{document}

\preprint{APS/123-QED}

\title{\boldmath Constraining the SMEFT with a differential cross section measurement of \tWZ production at the HL-LHC.}

\author{James Keaveney}
 \email{james.keaveney@uct.ac.za}
\affiliation{%
 University of Cape Town,\\ University Avenue, Cape Town, South Africa
}%


\begin{abstract}
A prospective measurement of the differential cross section of \tWZ production with respect to the transverse momentum of the \PZ boson using a general-purpose detector at the High-Luminosity Large Hadron Collider (HL-LHC) is described. The response of a general-purpose detector at the HL-LHC is simulated and used to estimate the uncertainties and covariances of the differential cross section measurement. Constraints on the Standard Model Effective Field Theory (SMEFT) enabled by the measurement are estimated. A parametric model of the differential cross section in the SMEFT is constructed and is used to determine the expected posterior probability function of six SMEFT Wilson coefficients and the expected 95\% Bayesian credible intervals for each coefficient and pair of coefficients. The intervals suggest that for all coefficients, the measurement will provide competitive but weaker constraints than those derived from other \mbox{HL-LHC} measurements involving top quarks and \PZ bosons. However, as the measurement is simultaneously sensitive to to a unique set of SMEFT coefficients, it will provide a useful input to a global SMEFT analysis that considers many operators.
\end{abstract}

\maketitle

\section{Introduction}
The relevance of the top quark's couplings to theories of new physics is well documented \cite{Bezrukov:2014ina,Beneke:2000hk,Tait:2000sh, Georgi:1994ha}. Given the lack of evidence of resonant production of new particles in the Large Hadron Collider (LHC) data, attention has turned to new physics scenarios in which new particles have masses around an energy scale $\Lambda$ that is large with respect to the scales directly probed at the LHC. The Standard Model Effective Field Theory (SMEFT) is a theoretical framework that describes the effects of such heavy new particles on lower energy observables. The SMEFT Lagrangian is obtained by extending the SM Lagrangian with higher-order operators \cite{Grzadkowski:2010es}. In the work presented here, only the dominant, dimension-6 operators are considered with the contribution of each operator to the Lagrangian scaled by a dimensionless Wilson coefficient $c_{i}$ divided by $\Lambda^{2}$. If the scale of new physics is indeed large with respect to the LHC energies, significant deviations of measurements of $c_{i}$ from the SM expectation ($c_{i}=0$) may be the first evidence of new physics. As the particle physics community prepares for the HL-LHC where proton collision datasets of unprecedented size will be produced, it is crucial to identify previously unexplored measurements that have the potential to improve the precision at which \ci can be determined. In this document, the potential of a measurement of one such observable, the differential cross section of \tWZ production, at a general-purpose detector is explored.

\section{Expectations for the \tWZ process at the HL-LHC}
\label{sec:twz}
The \tWZ process refers to the electroweak production of a single top quark in association with a \PW boson and a \PZ boson.  In 13 TeV proton collisions, \tWZ is predicted to have a SM cross section of $\approx 107$ fb at NLO in QCD when an operational definition of the \tWZ process is adopted \cite{Faham:2021zet}. The modelling of this process at NLO in QCD has also been explored in~\cite{Bylund:2016qau} and the sensitivity of the inclusive \tWZ cross section to the SMEFT coefficients has been quantified in~\cite{Faham:2021zet, Maltoni:2019aot, Mimasu:2021cpd}. As the \PZ boson may be radiated from the initial-state bottom quark, the final-state top quark, or the final-state \PW boson, the process already exhibits complex phenomenology at LO in the SM by simultaneously embedding the \Pqt-\PZ, \Pqb-\PZ, and \PW-\PZ electroweak couplings. Consequently, the \tWZ process is sensitive to multiple SMEFT operators. These operators modify SM interaction vertices and generate new vertices not present in the SM. Not all of the operators that affect the \tWZ process involve the top quark field. Although the primary goal of this work is to estimate potential constraints on SMEFT coefficients related to top quark couplings, ignoring all other operators that affect the \tWZ process could lead to an overestimation of the precision of the constraints. Therefore the following set of operators are chosen on the basis of them having a significant effect on either the overall rate or kinematics of the \tWZ process. However the list is not exhaustive and investigation of the expected complex interplay within a larger set of operators in an NLO cross section calculation is left to future work. Similarly, interference effects at NLO between the \tWZ and \ttZ processes may be important to a future analysis of \tWZ measurements at the HL-LHC in the SMEFT. In this paper, the implementation of the SMEFT in the \smeftatnlo package \cite{Degrande:2020evl} is adopted. The operator definitions in \smeftatnlo are provided explicitly in \cite{Degrande:2020evl_smeftatnlo}. The definitions of first three of these operators, taken from \cite{Degrande:2020evl_smeftatnlo}, are:

\begin{itemize}
    \item[] $\mathcal{O}^{(3)}_{\phi{}\Pq{}}~=~i(\phi^{\dagger} \overset{\text{\tiny$\bm\leftrightarrow$}}{D}_{\mu} \tau_{I} \phi) (\bar{Q}\gamma^{\mu} \tau^{I} Q)$
    \item[] $\mathcal{O_{\Pqt{}\PW}}~=~i(\bar{Q} \tau^{\mu \nu} \tau_{I} t)\tilde{\phi} W^{I}_{\mu\nu}$+h.c.
    \item[] $\mathcal{O}_{\Pqt{}G} = i g_{s} (\bar{Q} \tau^{\mu \nu} T_{A} t)\tilde{\phi} G^{A}_{\mu \nu}$ + h.c.
\end{itemize}

A linear combination of the effects of the operators $\mathcal{O_{\Pqt{}\PW}}$ and $\mathcal{O_{\Pqt{}B}}~=~i(\bar{Q} \tau^{\mu\nu} t)\tilde{\phi} B_{\mu\nu}$+h.c. according to to the Weinberg angle is also considered leading to four independent degrees of freedom in the SMEFT analysis.

The advantages of \tWZ measurements in constraining the SMEFT coefficients are detailed in \cite{Mimasu:2021cpd}. The effect of the SMEFT operators on the \tWZ process show an energy dependence that is more pronounced than that of alternative processes such as \tZq. Thus as more data is recorded and measurements of higher-energy phase space of \tWZ become possible, constraints on SMEFT coefficients from \tWZ will continue to improve after constraints from processes with weaker energy dependence become saturated. This serves as particular motivation to assess the potential of measuring \dsigptz as a probe of the high-energy phase space of \tWZ. 

 Only two final-state topologies of the \tWZ process will be measurable with reasonable precision at the HL-LHC. The first topology ($3\ell$) corresponds to the case where the \PZ boson decays to a pair of oppositely-charged electrons or muons and at least one other electron or muon is produced from the decays of the \PW bosons. The second topology ($4\ell$) corresponds to the case where the \PZ boson and both \PW  bosons decay into electrons or muons. 
 
 The estimation of expected experimental uncertainties and inter-bin covariances in the measurement of \dsigptz at the HL-LHC is now described. For all of the following studies, a proton-proton centre-of-mass energy of 13 TeV and an integrated luminosity of 3000 fb$^{-1}$ is assumed. A set of Monte-Carlo simulation event samples corresponding to the SM expectation for the signal and background processes were generated using \mg \mgversion for the matrix-element calculation and \pythia \pythiaversion for the parton shower and hadronisation modelling. For the $3\ell$ topology the considered backgrounds are \ttZ, the production of a \PW and \PZ boson in association with at least two jets including b-jets (\WZbb), and the production of the a single top quark in association with a \PZ boson (\tZq). For the $4\ell$ topology the considered backgrounds are \ttZ, and the production of a pair of \PZ bosons in association with at least two least two jets not including b-jets (\ZZjj). Other background processes are assumed to be negligible in the context of a \tWZ cross section measurement. 
 
\begin{figure}[]
\centering
\includegraphics[width=.45\textwidth]{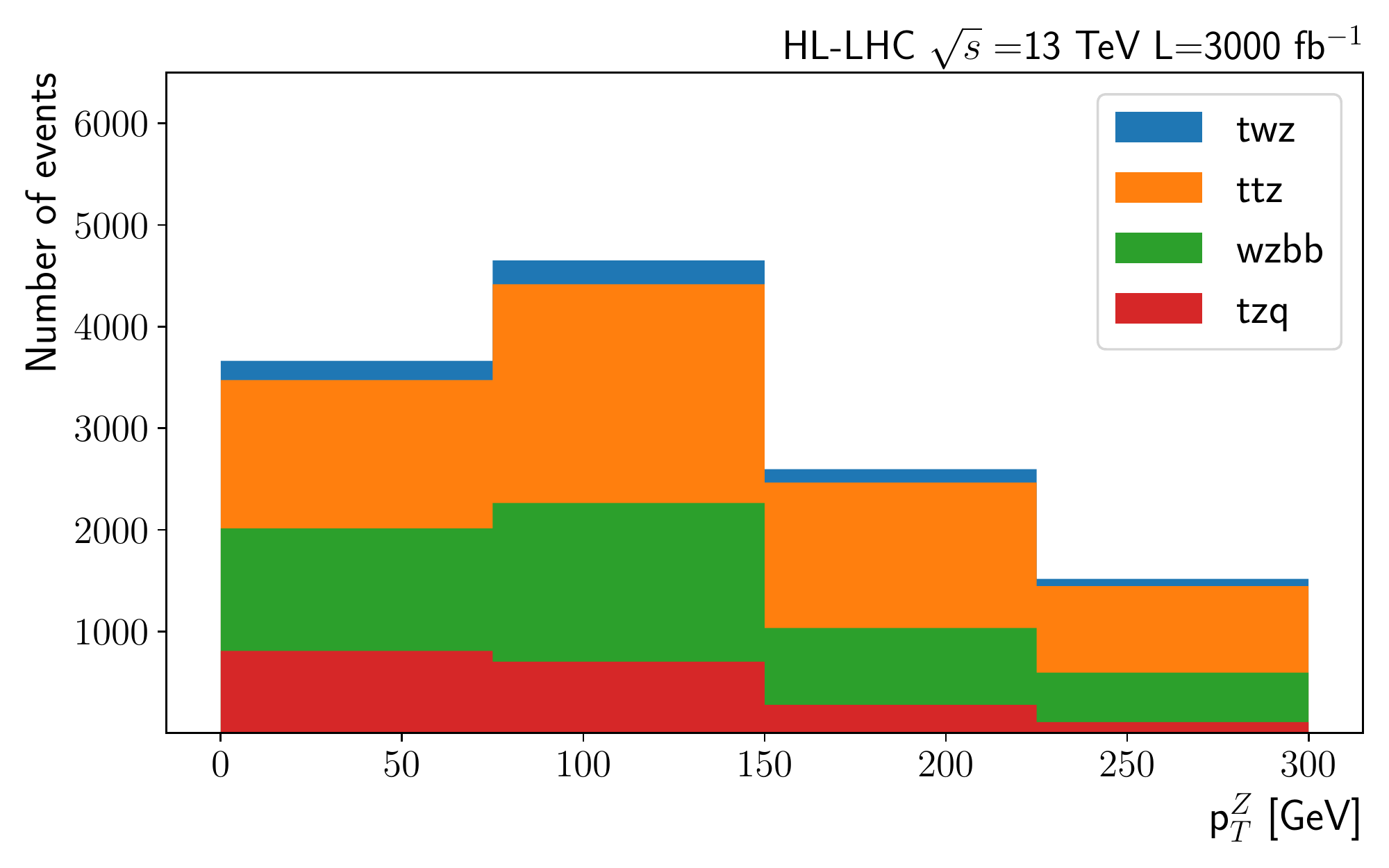}
\caption{The expected distributions of \ptz for the signal and background processes after the application of the $3\ell$ selection at the HL-LHC are shown.}
\label{fig:3l_expected}
\end{figure}

\begin{figure}[]
\centering
\includegraphics[width=.45\textwidth]{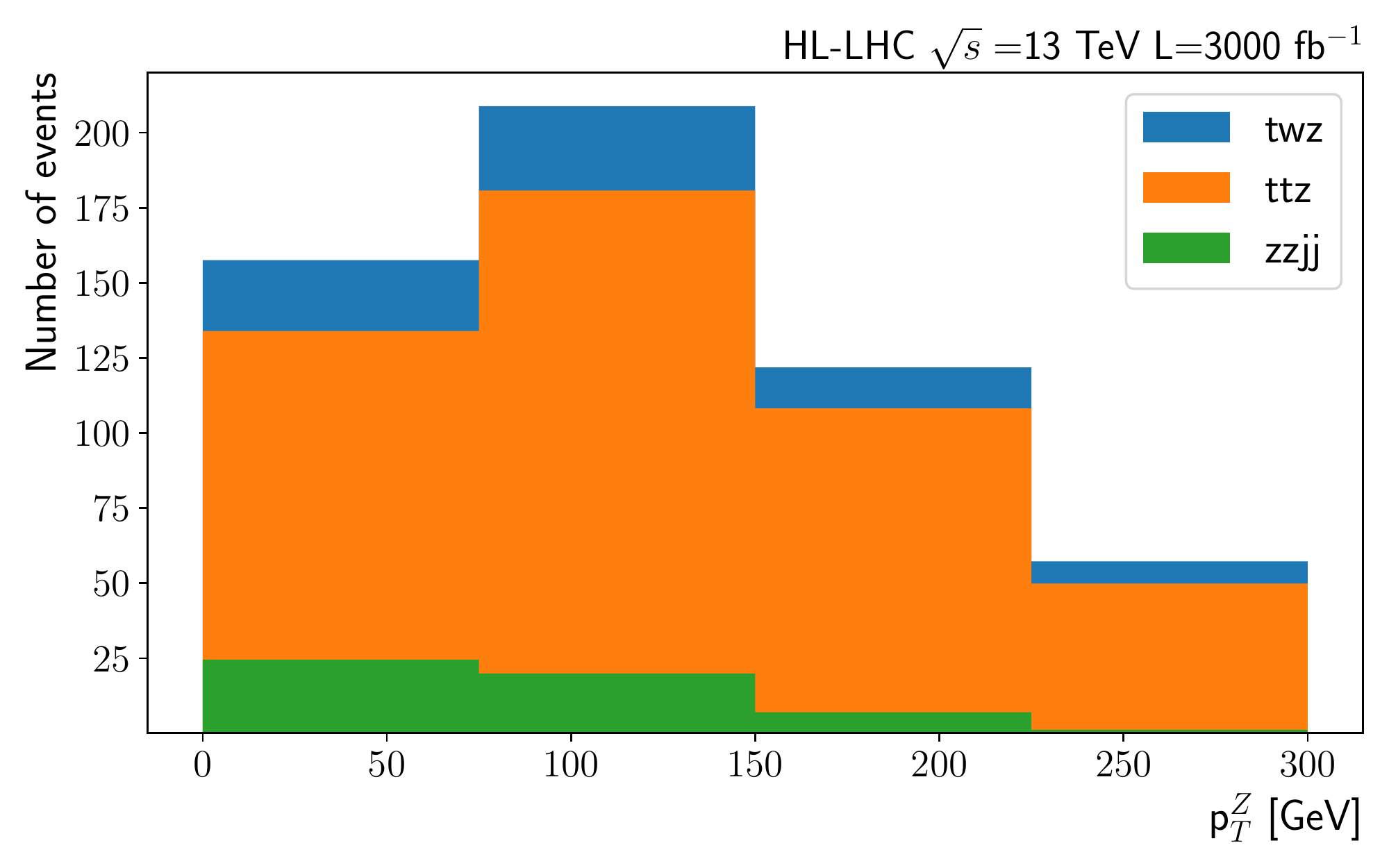}
\caption{The expected distributions of \ptz for the signal and background processes after the application of the $4\ell$ selection at the HL-LHC are shown.}
\label{fig:4l_expected}
\end{figure}

 The samples were processed with the \delphes \delphesversion framework \cite{deFavereau:2013fsa} configured to simulate the response of a general-purpose detector in the experimental conditions of the HL-LHC. Two mutually-exclusive event selection schemes were applied that target the two topologies while suppressing the respective background processes. In Table \ref{tab:sel_scheme}, the common criteria applied to the \emph{objects}, i.e, electrons, muons and jets, reconstructed within each event and the two sets of criteria to select events for the two topologies are detailed. In Figures \ref{fig:3l_expected}, and \ref{fig:4l_expected}, the distributions of the transverse momentum of the \PZ boson candidate reconstructed in each event for the two topologies respectively are shown. The distributions are \emph{stacked} such that for each figure the sum of the illustrated components corresponds to the total expected distribution. In the case of the signal and \ttZ processes, the expected distributions are normalised according to the NLO QCD cross sections. For the other background processes the LO QCD cross sections are used.
 
 The figures indicate that the \tWZ signal is small compared to the backgrounds. Hence the total uncertainties on the measurement of \dsigptz will be dominated by the uncertainties on the background contributions. In a recent CMS paper\cite{CMS:2021ugv}, the differential cross sections of the \tZq process and background contributions were estimated via a maximum-likelihood fit. With this technique the uncertainty on the background contribution is largely determined by the statistical uncertainties on the background contributions and additional modelling and instrumental uncertainties. Assuming that such a technique will be applied to the measurement of \dsigptz, the total expected uncertainty on the measurement is estimated as the quadrature sum of the statistical uncertainty on the total background and a 10\% systematic uncertainty on the measured cross section to account for additional modelling effects. However there is significant potential for further reduction of the background contributions and hence greater precision on the \dsigptz measurement. The \WZbb and \tZq backgrounds in the $3\ell$ topology could be reduced via the reconstruction and selection of hadronically-decaying \PW bosons from dijet systems that are a feature of the signal process but will not be present in the \WZbb process when the ($3\ell$) topology is selected. Due to the fundamental similarity between the \tWZ and \ttZ processes, suppression of the of \ttZ background is more difficult. However multivariate techniques can exploit subtle differences in the kinematics of the \tWZ and \ttZ processes associated with the differing number of resonant top quarks. This approach has been successfully employed in the measurement of the differential cross sections of the \tZq process in \cite{CMS:2021ugv}. The development of multivariate algorithms depends crucially on precise and calibrated simulation of the detector response and is thus not attempted here in this exploratory study.
 
 The statistical uncertainties within each bin depend on the binning scheme. Increasing the number of bins will yield greater sensitivity to the SMEFT coefficients as the shape of the distributions is measured in greater detail. Furthermore the number of coefficients that can be independently constrained by a single absolute differential cross section is limited by the number of bins. Conversely, statistical uncertainties in each bin increase as the number of bins increases, degrading the constraints on the SMEFT coefficients. A binning scheme that consisting of four equal-width bins between 0 and 300 GeV that aims to balance these concerns is adopted. 

\begin{figure}[]
\centering
\includegraphics[width=.48\textwidth]{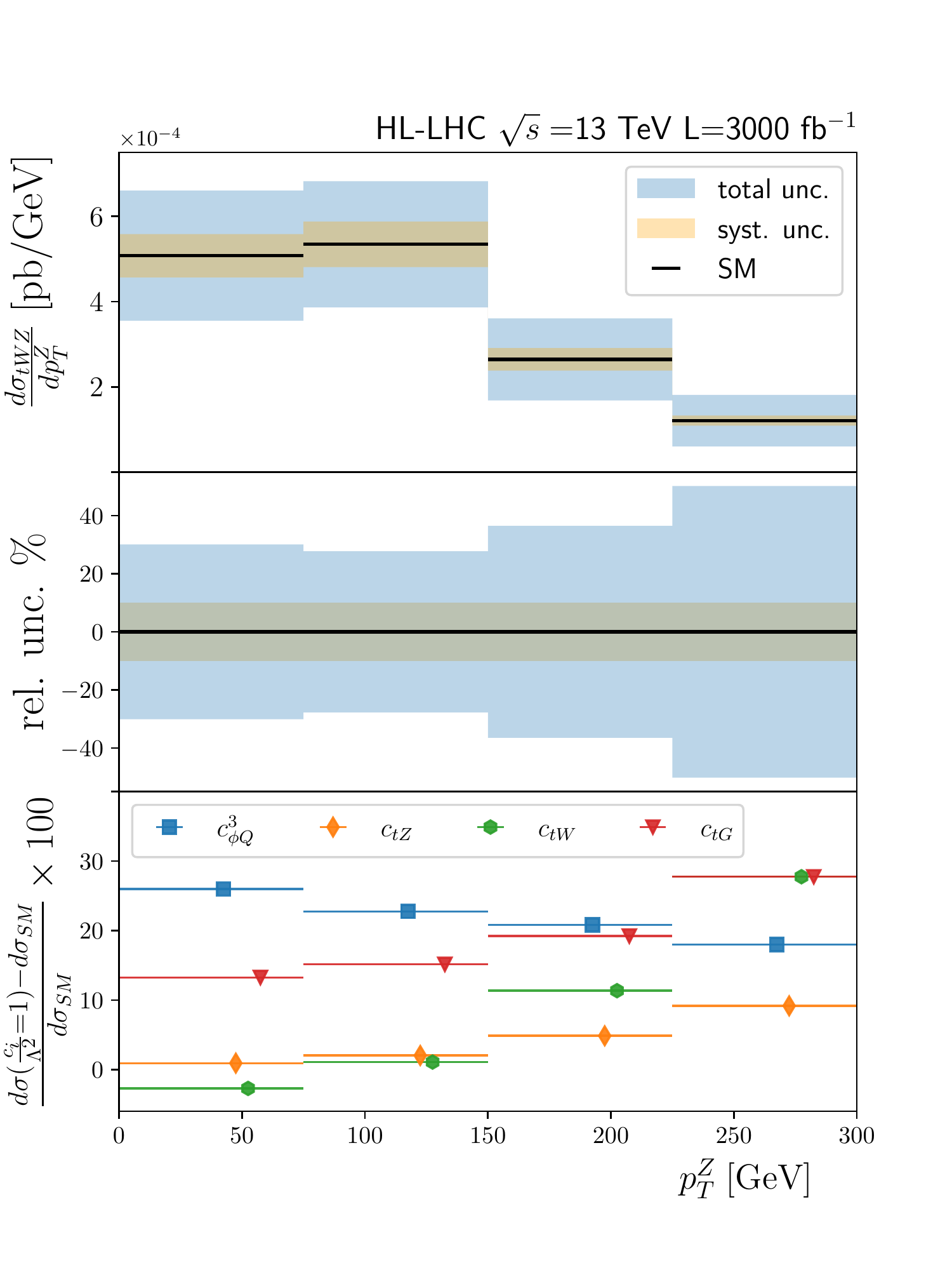}
\caption{The prediction for \dsigptz at 13 TeV calculated in the SM with \mg  at LO in QCD but normalised according to the NLO cross section is shown in the upper panel. The expected total and systematic uncertainties of a measurement of this distribution at the HL-LHC are shown by the blue and gold bands respectively. The central panel shows the uncertainties expressed as percentages of the central values of the prediction. The lower panel shows the differences between the SM prediction and a set of SMEFT predictions corresponding to each coefficient set to a value of $\ci = 1.0$ TeV$^{-1}$ with all other coefficients set to 0 TeV$^{-1}$. The differences are expressed as percentages of the SM prediction.}
\label{fig:expected_measurement}
\end{figure}

 Differential cross sections are typically measured after an unfolding procedure in which the resolution effects of the detector response are accounted for. The unfolding procedure introduces statistical correlations between the measured values of \dsigptz in the bins of the unfolded differential cross section due to migrations of events across bin boundaries. These migrations are modelled with a migration matrix that encodes the probabilities of events in a given bin at unfolded level being observed in each bin at detector level. Similarly, sources of systematic uncertainty that cause correlated effects in multiple bins introduce additional inter-bin correlations. The uncertainties on a measurement of the unfolded \dsigptz are estimated by combining the expected detector-level distributions in the two topologies and unfolding the combined distributions using the \roounfold package \cite{Prosper:2011zz} in which the Iterative Bayesian unfolding algorithm \cite{DAgostini:1994fjx} is applied. The number of iterations for this algorithm was set to 4 as this was found to produce good agreement between the true and unfolded \dsigptz distributions. The robustness of the unfolding setup in the presence of SMEFT signals was verified by applying an multiplicative factor to \dsigptz at detector- and truth-levels that depended linearly on \ptz and varied from 1.0 in the first bin to 1.5 in the last bin and applying the unfolding that was developed with the SM \tWZ prediction. The agreement between unfolded and true \dsigptz distributions remained excellent in the presence of this signal.  The binning of the unfolded distribution is chosen to match that of the detector-level distributions. The migration matrix is estimated by comparing the simulated sample of \tWZ events at detector- and generator-levels.
In the upper panel of Figure \ref{fig:expected_measurement}, the SM prediction for \dsigptz in the chosen binning scheme is shown. The estimated total uncertainty on after the application of the unfolding process is shown by the blue band. The estimated systematic uncertainty is shown by the gold band. The central panel of the figure shows the same two uncertainties expressed as percentages of the SM expectation. As none of the suggested kinematic reconstruction or multivariate techniques have been applied, these estimates of the uncertainties may be considered conservative. 

\begin{table}[!htb]
\def\arraystretch{1.3}
\caption{The criteria applied to objects and events for the two topologies are shown. Where the object criterion for the two topologies differ, the criterion applied in the $4\ell$ topology is shown in brackets.}
\label{tab:sel_scheme}
\begin{tabular}{|c|c|}
 \hline
 \multicolumn{2}{|c|}{\bf{Objects}} \\
 \hline
 \bf{Object type} & \bf{Criteria}  \\
 \hline
 $\ell$ & \pt $>$ 15 (10) GeV, iso $<$ 0.1, $|\eta| < 4$ \\
 \hline
 \PZ candidate & $60 < m_{\ell,\bar{\ell}} < 120 $ GeV \\
 \hline
  jet & \pt $> $ 30 (20) GeV, $|\eta | < 4.5$ \\
 \hline
 \multicolumn{2}{|c|}{\bf{Events}} \\
 \hline
 \bf{Topology} & \bf{Criteria} \\
 \hline
 $3\ell$ & $N_{\ell} = 3$, $N_{Z} = 1$, $N_{jet} \geq 3$, $N_{btag} \geq 1$ \\
  \hline
 $4\ell$ & $N_{\ell} = 4$, $N_{Z} = 1$, $N_{jet} \geq 1$, $N_{btag} \geq 1$ \\
 \hline
\end{tabular}
\end{table}

\section{Differential cross sections in the SMEFT}
\label{sec:diff_xsec}
If the SMEFT Lagrangian contains $n$ dimension-6 operators $c_i$ and the number of such operators in each Feynman diagram is limited to one, any cross section in the SMEFT can be expressed as a 2$^{nd}$ order multivariate polynomial in $c_{i}$:

\begin{equation*}
    \sigma( c_{i} ) = \sigma_{SM} + \sum^{i=n}_{i=1} \frac{c_{i}} {\Lambda^{2}}\beta_{i} + \sum^{j=n}_{j=1} \sum^{i=n}_{i=1} \frac{c_{i}c_{j}} {\Lambda^{4}}\beta_{ij} 
\end{equation*}

In this expression, $ \sigma( c_{i} ) $ is the cross section in the SMEFT, written explicitly as a function of the $n$ SMEFT coefficients and $\sigma_{SM}$ is the SM cross section. The $\frac{c_{i}} {\Lambda^{2}}\beta_{i}$ terms represent the contributions from the product of the SM amplitude and the amplitudes of diagrams containing a dimension-6 vertex. The  group of terms $\frac{c_{i}c_{j}} {\Lambda^{4}}\beta_{ij}$ represent the $n^{2}$ contributions from the products of amplitudes of diagrams containing a vertex from the $i^{th}$ operator with the amplitudes of diagrams containing a vertex from the $j^{th}$ operator and thus includes the contributions from the squares of the amplitudes of diagrams containing a vertex from the $i^{th}$ operator. As $\frac{c_{i}c_{j}} {\Lambda^{4}}\beta_{ij}  = \frac{c_{j}c_{i}} {\Lambda^{4}}\beta_{ji}$, only $(\frac{(n)(n+1)}{2})$ independent contributions are present in the $\sum^{j=n}_{j=1} \sum^{i=n}_{i=1} \frac{c_{i}c_{j}} {\Lambda^{4}}\beta_{ij} $ term. The terms $\sigma_{SM}$, $\beta_{i,1}$ and $\beta_{i,j}$, are specific to the process and represent the $b$ unknowns of the polynomial where $b = 1 + n + (\frac{(n)(n+1)}{2})$. Thus for $n$ operators, $b$ is the $(n+1)^{th}$ triangular number. If the $\sigma$ and $\beta$ terms are vectors with elements corresponding to the contributions to the differential cross section within each bin, then the polynomial represents a parametric model of a differential cross section.

An exact solution for the unknowns of the polynomial can be found by making a minimum number of theoretical calculations for $\sigma_{SMEFT}$ at distinct points in $c_{i}$ space and forming a system of linear equations in an approach known as \emph{morphing} \cite{TheATLAScollaboration:2015bhf, Baak:2014fta}. An alternative approach based on Bayesian reasoning is presented in \cite{Castro:2016jjv}. In general, Monte-Carlo based theoretical predictions carry significant statistical uncertainties, especially in regions of phase space with small cross sections. Furthermore, in SMEFT calculations involving multiple operators, one or more of the unknowns may be extremely small relative to the others at all of the points in $c_{i}$ space used to construct the system of linear equations. Hence these relatively small unknowns can be imprecisely determined by the exact solution to the linear system. When contributions associated to these unknowns become large in other regions of the SMEFT coefficient space, inaccuracies in the predictions of the morphing model can arise. The approach pursued here, termed \emph{regression morphing}, alleviates this issue. Instead of producing an exact solution of a system of linear equations, the unknowns are determined by minimising the sum of squared differences $S$ between $p$ theoretical predictions for the differential cross section at randomly distributed points in the SMEFT coefficient space and the corresponding predictions of the regression morphing model. For a model derived from $p$ predictions, each with $q$ bins, $S$ is defined as 
\begin{equation*}
S(\vec{\beta}) = \sum^{i=p}_{i=1} \sum^{j=q}_{j=1} (d \sigma_{i,j} - {x(\vec{\beta})_{i,j}} )^{2}    
\end{equation*}
where $S$ is written explicitly as a function of a vector of the model's unknowns $\vec{\beta}$, $d \sigma_{i,j}$ is the differential cross section in the $j^{th}$ bin of the $i^{th}$ theoretical prediction and $x(\vec{\beta})_{i,j}$ is the corresponding prediction of the model. The regression morphing model can be made arbitrarily accurate in any region of the SMEFT coefficient space by including a sufficient number of predictions when minimising $S$. 

\section{Constraining SMEFT coefficients}
\label{sec:cons_twz}
To construct a regression morphing model for \dsigptz at the HL-LHC, 270 \emph{training} predictions for \dsigptz 13~TeV proton collisions were generated at random points in the \ci space. The predictions were produced with the \smeftatnlo \cite{Degrande:2020evl} package implemented in the \mg Monte Carlo generator version \mgversion \cite{Frederix:2018nkq} at LO in QCD and in fixed-order mode. Although fixed-order predictions lack the parton shower and hadronisation modelling usually important to the modelling of hadron collider observables, the speed of generation and lack of a need to process large MC event samples makes them invaluable in the construction of regression morphing models and are sufficient to demonstrate the sensitivity of \dsigptz to the SMEFT coefficients. Similarly, the usage of NLO predictions would yield more precise estimates of \dsigptz and expected constraints on the \ci. However, NLO predictions would require either the application of a diagram removal scheme to suppress components of the \tWZ calculation that overlap with the \ttZ process or a calculation that fully accounts for overlapping components and associated interference effects. Such a theoretical treatment is left to future work.

To validate the model, 30 \emph{test} predictions, statistically independent from the training predictions, were generated at random points in the SMEFT coefficient space and compared to the corresponding predictions of the model. The binned residuals between the test predictions and model predictions expressed as percentages of the test predictions are Gaussian-distributed with a mean of -0.0005 \% and standard deviation of 0.0089 \%, demonstrating that the inaccuracies in the regression morphing model model are negligible. The posterior probability density function \postfunc is used to derive constraints on the four SMEFT coefficients where $c_{i}$ are the SMEFT coefficients, $x$ are expected measured values of \dsigptz in the SM scenario, and $\Sigma$ is the estimated covariance matrix of $x$ given the previously stated assumptions on uncertainties and correlations.
\begin{figure*}[]
\centering 
\includegraphics[width=.83\textwidth]{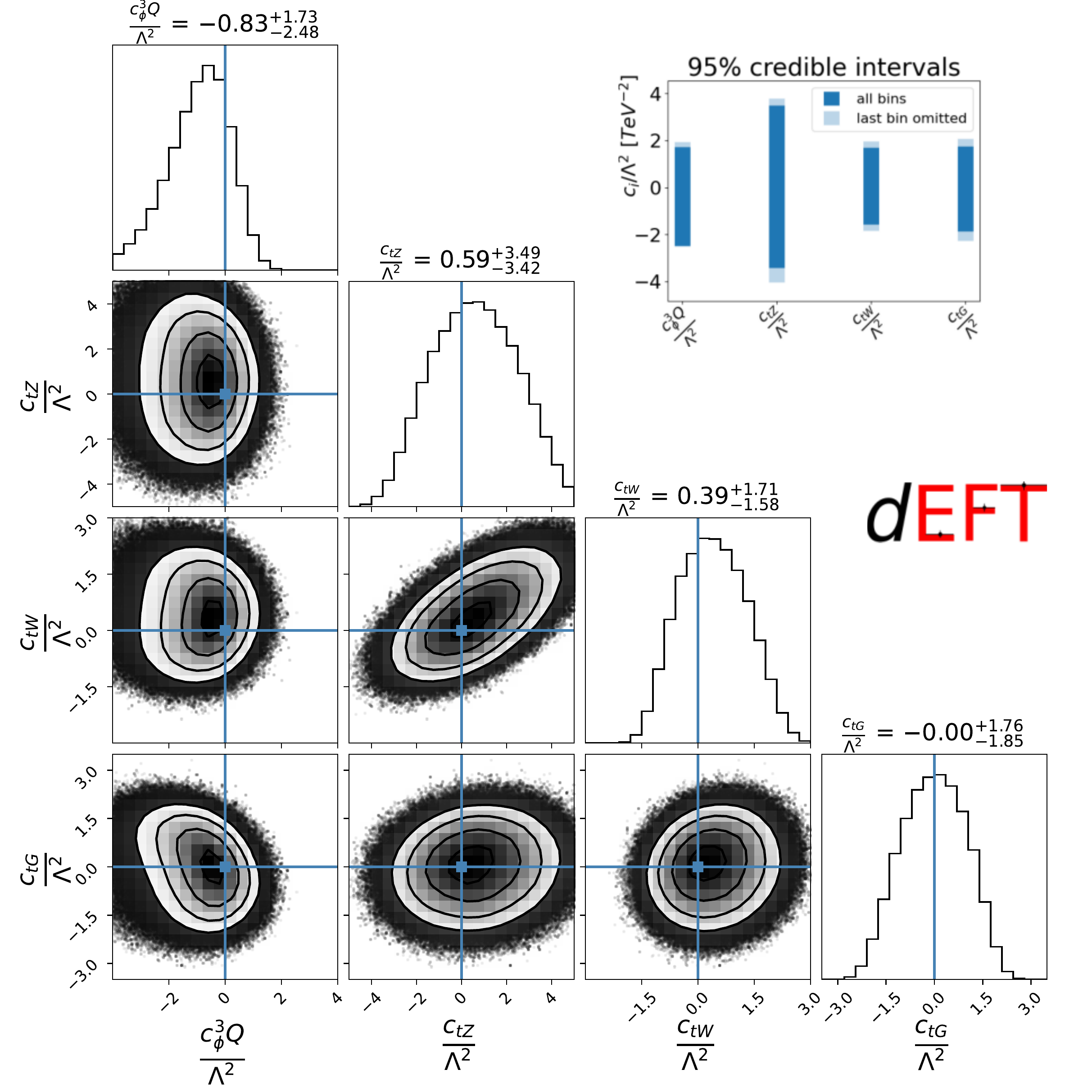}
\caption{An array of the marginalised posterior probability density functions is shown. The 1-D functions for each of the SMEFT coefficients are placed along the grid diagonal. For each 1-D function, the median $c_{i}$ value is indicated with the difference between this value and the limits of the 95\% credible intervals given by the superscripts and subscripts and also by the vertical dashed lines. The 95\% credible intervals are also represented by the blue bars in the panel on the top right after they have been shifted such that the best-fit values equal 0. The 2-D functions for each pair of coefficients are shown in the other grid spaces. The white regions on the 2-D functions are CIs constructed to coincide with the 1-D CIs.}
\label{fig:constraints}
\end{figure*}

The expected data are assumed to be Gaussian-distributed hence the log-likelihood is approximated as a $\chi^{2}$ function accounting for the inter-bin correlations of the data introduced by the \emph{unfolding} process used to correct for detector resolution effects. This leads to the final expression for \postfunc:
\begin{equation*}
\postfunc = \priorfunc (-0.5)\likelihood 
\end{equation*}
where \priorfunc is the prior probability of the SMEFT coefficients, $d(\frac{c_{i}} {\Lambda^{2}}, x)$ is a column vector of the residuals between the regression morphing model and expected data for the coefficient values $\frac{c_{i}} {\Lambda^{2}}$, $d(\frac{c_{i}} {\Lambda^{2}}, x)^{T}$ is the transpose of $d(\frac{c_{i}} {\Lambda^{2}},x)$, and $\Sigma^{-1}$ is the inverse of the covariance matrix of the expected data. This expression ignores the constant factor present in the canonical definition of the posterior probability which has no effect on the constraints. The prior probability is chosen to be constant as a function of all SMEFT coefficients within an allowed region that is much larger than the expected 95\% credible intervals (CI) of the posterior function and zero elsewhere. The central values of the expected data are chosen the match the prediction of the model in the SM ($c_i=0$) hypothesis. A numerical estimation of \postfunc is produced using the \mc package \cite{Foreman-Mackey:2013} that implements a Markov Chain Monte Carlo algorithm. The constraint on a given single SMEFT coefficient is defined as the 95\% CI of the 1-D pdf obtained by marginalising \postfunc over the other coefficients centered around 0. In order to verify the validity of the SMEFT in the energy regime of a differential cross section it has been suggested to compare results with and without the highest energy bin included \cite{Aguilar-Saavedra:2018ksv}. Thus the analysis is also performed after the removal of the last bin of the \dsigptz distribution and hence the last row and column of the covariance matrix. The difference in results between these two setups is small indicating that the sensitivity of the \dsigptz measurement is not dominated by the last bin and checks of the validity of the SMEFT model in this energy regime will be possible in an analysis using real HL-LHC data.

In Figure \ref{fig:constraints}, the 1- and 2-D marginalised posterior functions are shown in a grid arrangement. The marginalisations of \postfunc are performed with the \corner package \cite{Foreman-Mackey:2016}. The median values of the posterior function and the positive and negative distances to the edges of the CI are shown in text above each 1-D pdf. The edges of 95\% CI centered around the median for the each of the six 1-D function are illustrated by vertical dashed lines. The blue lines indicate the SM expectation ($c_{i}$). The white regions on the 2-D distributions are CIs constructed as to coincide with the 1-D CIs. The SM expectations for the $c_i$ are indicated by intersections of the blue lines. An additional plot with the final constraints and those produced with the last bin omitted is provided in the upper-right corner of the figure. The sensitivity of \dsigptz to all six coefficients is apparent by the finite widths of the CI. Correlations between the values of $c_{t\PZ}$ and $c_{tW}$ are apparent in the shape of the corresponding 2-D functions. These correlations could be mitigated by utilising an alternative differential distribution or by including measurements of other processes such as \ttZ and \tZq that are sensitive to subsets of these operators. 

In Figure \ref{fig:compare_constraints}, the constraints are shown alongside frequentist 95\% confidence intervals on subsets of these four coefficients reported a set of five other analyses for the purposes of broad comparison. The set comprises an analysis of an array of top quark measurements using LHC data with the \smefit framework \cite{Hartland:2019bjb}, a similar SMEFT analysis focusing top quark data from the LHC performed by the \ifit collaboration \cite{Miralles:2021dyw}, constraints from a measurement of the differential cross section at 13 TeV from the CMS experiment \cite{CMS:2019too},  expected constraints from a measurement of the inclusive \ttZ cross section at the HL-LHC by the CMS experiment \cite{CMS-PAS-FTR-18-036} and an estimation of the constraints obtainable from a future measurement of the \ttZ cross section at the HL-LHC in the channel where the \PZ boson decays to a pair of neutrinos \cite{HajiRaissi:2020eob}. Due to the LO modelling of the \tWZ signal, the approximations made to derive expected uncertainties and covariances of the \dsigptz measurement, and the different statistical definitions of the constraints, quantitative comparison of the constraints is difficult. 
However it is apparent that the constraint on \ctZ derived from the measurement of \dsigptz will be competitive with those already obtained from global analyses of LHC data using the \smefit framework \cite{Hartland:2019bjb}. The constraints on \ctZ \dsigptz are significantly weaker than derived from measurements of the \ttZ process. However it is not clear that this difference will remain when the many four-quark operators that affect the \ttZ process but not the \tWZ process are considered.  While the constraints on the other \ci appear significantly weaker than those from other measurements, the large number of operators affecting \dsigptz suggests that this observable will be useful in improving constraints in analysis simultaneously utilising large sets of top quark measurements. 

The construction and validation of the regression morphing model as well as interfaces to \mc and \corner is provided by \deft, a python package available on GitHub\footnote{The \deft codebase is available under a GNU GENERAL PUBLIC LICENSE at \deftlink.}.

\begin{figure*}[]
\centering 
\includegraphics[width=0.95\textwidth]{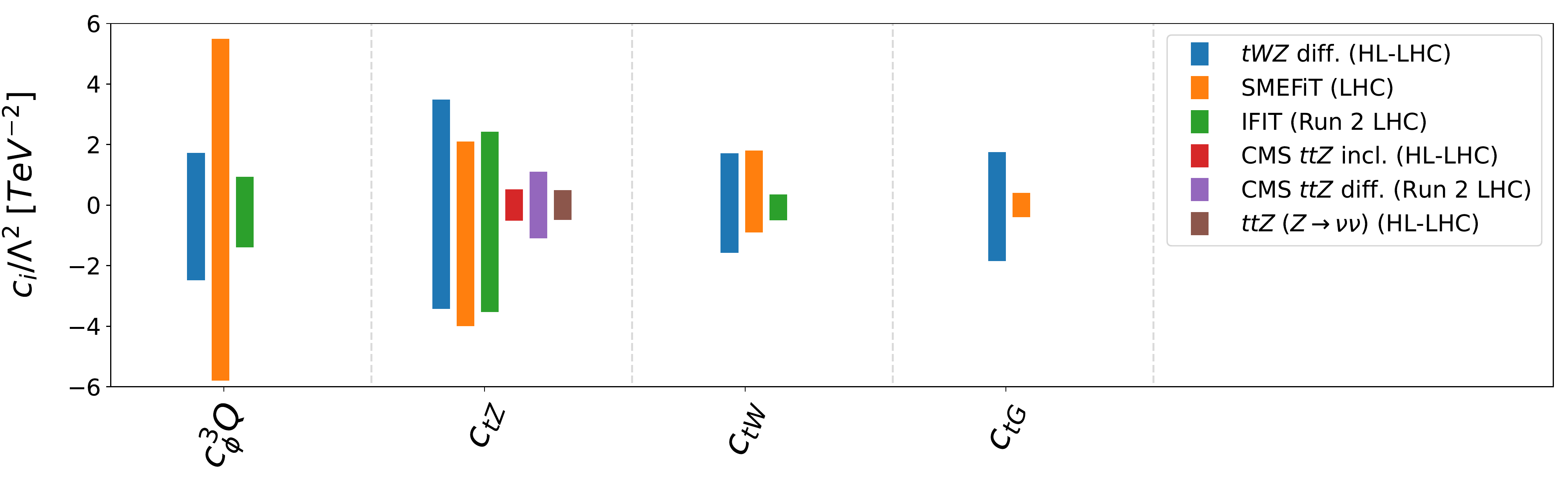}
\caption{The expected 1-D 95\% CI on the four SMEFT coefficients produced in this work are shown by the blue bars. Frequentist 95\% confidence intervals on subsets of these coefficients reported elsewhere are shown for broad comparison \cite{Hartland:2019bjb,Miralles:2021dyw,CMS:2019too,CMS-PAS-FTR-18-036, HajiRaissi:2020eob}. }
\label{fig:compare_constraints}
\end{figure*}

\section{Conclusion and outlook}
An analysis of the effects of a set of four dimension-6 operators in the SMEFT on the differential cross section of \tWZ production with respect to the transverse momentum of the \PZ boson is presented. Estimates of the uncertainties and covariances of the a measurement based on LHC data are used to estimate the expected constraints on the SMEFT coefficients. A method termed regression morphing is used to construct a parametric model of \dsigptz as a function of the SMEFT coefficients. 

A numerical approximation of the expected posterior probability function is derived. Constraints on the coefficients are defined as 95\% credible intervals determined by marginalising the posterior probability function.
In the case of $c^{(3)}_{\phi{}\Pq{}}$, the constraint is comparable or stronger than those derived from arrays of LHC measurements \cite{Hartland:2019bjb,Miralles:2021dyw} and is comparable to expected constraints from measurements of the inclusive \ttZ cross section at the HL-LHC and the differential \ttZ cross section at the LHC \cite{Hartland:2019bjb,Miralles:2021dyw,CMS:2019too,CMS-PAS-FTR-18-036, HajiRaissi:2020eob}. 

The measurement will also constrain the other coefficients to a lesser degree than other HL-LHC measurements and LHC measurements considered. Hence we conclude that measurements of \dsigptz at the HL-LHC will be an important input to a global analysis of SMEFT coefficients with the potential to increase the discovery potential for new physics in the top quark sector in the HL-LHC data.

\begin{acknowledgments}
The author would like to acknowledge the usage of the High Performance Computing cluster at the University of Cape Town and to thank Ilaria Brivio and Stephen Farry for much useful feedback.
\end{acknowledgments}

\end{document}
%